\documentclass[a4paper,twoside]{article}
\newcommand{\affil}[1]{$^{\rm #1}$}
\usepackage[authoryear]{natbib}
\bibpunct{(}{)}{;}{a}{}{,}
\usepackage{graphicx}
\usepackage{lscape}
\usepackage{color}
\date{} 
\title{\large\bf\flushleft Interpretation of CEMP(s) and
CEMP(s+r) stars with AGB models}
\author{\parbox{\textwidth}{\flushleft
\vspace{-0.5cm}
{\it Sara Bisterzo\affil{A,E}, Roberto Gallino\affil{A}, Oscar Straniero\affil{B},
Wako Aoki\affil{C,D}}\\
\vspace{0.4cm}
{\small \affil{A}\,Dipartimento di Fisica Generale, Universit\`a di Torino, via P. Giuria, 1, 10125 Torino, Italy}\\
{\small \affil{B}\,INAF Osservatorio Astronomico di Collurania, via M. Maggini, 64100 Teramo, Italy}\\
{\small \affil{C}\,National Astronomical Observatory of Japan, 2-1-21 Osawa, Mitaka, Tokyo, 181-8588 Japan}\\
{\small \affil{D}\,Department of Astronomical Science, The Graduate University of Advanced
Studies, Mitaka, Tokyo, 181-8588 Japan}\\
{\small \affil{E}\,Email: bisterzo@ph.unito.it}}}
\begin{document}
\twocolumn[
\begin{changemargin}{.8cm}{.5cm}
\begin{minipage}{.9\textwidth}
\vspace{-1cm}
\maketitle
%
%
\small{\bf Abstract:} 
Asymptotic Giant Branch (AGB) stars play a fundamental 
role in the $s-$process nucleosynthesis during their thermal 
pulsing phase.
The theoretical predictions obtained by AGB models at different
masses, $s-$process efficiencies, dilution factors and 
initial $r-$enrichment, are compared with spectroscopic observations of 
Carbon$-$Enhanced Metal$-$Poor stars enriched in $s-$process 
elements, CEMP(s), collected from the literature.
We discuss here five stars as example, 
CS 22880$-$074, CS 22942$-$019, CS 29526$-$110, HE 0202$-$2204, 
and LP 625$-$44.
All these objects lie on the main$-$sequence or on the giant 
phase, clearly before the TP$-$AGB stage: the 
hypothesis of mass transfer from an AGB companion, 
would explain the observed $s-$process 
enhancement. 
CS 29526$-$110 and LP 625$-$44 are CEMP(s+r) objects, and are interpreted 
assuming that the molecular cloud, from which the binary 
system formed, was already enriched in $r-$process elements by SNII pollution.
In several cases, the observed $s-$process distribution may be
accounted for AGB models of different initial masses with
proper $^{13}$C$-$pocket efficiency and dilution factor.
Na (and Mg), produced via the neutron capture chain starting from $^{22}$Ne, 
may provide an indicator of the initial AGB mass.


\medskip
\medskip
\end{minipage}
\end{changemargin}
]
\small

\section{Introduction}

A sample of about one hundred of
C$-$rich, $s-$rich metal$-$poor (CEMPs) and, when detected, lead$-$rich
stars have been observed in recent years
\citep[see][and references therein]{SCG08}.

Observed halo stars are of low mass ($M$ $\leq$ 0.9 $M_\odot$)
and long lifetimes, with effective temperatures
and surface gravities characteristic of main$-$sequence stars, subgiants
or giants, far from the Asymptotic Giant Branch 
(AGB) phase where the $s-$process is manufactured.
Therefore, the hypothesis of mass accretion of 
$s-$rich material from a more massive AGB companion, 
becomes essential to explain the overabundances 
detected in their spectra.
 
Our aim is to interpret the spectroscopic 
data of CEMP$-s$ stars with AGB 
theoretical models using different $^{13}$C$-$pocket 
efficiencies and initial masses.
Stellar model parameters have been derived over a set of 
AGB models obtained with the FRANEC code, as discussed 
in \citet{straniero03}.
Neutrons are released by the two reactions 
$^{13}$C($\alpha$, n)$^{16}$O and $^{22}$Ne($\alpha$, n)$^{25}$Mg.
The first reaction is the major neutron source.
When the H$-$shell is inactive, the so called Third
Dred-
ge$-$Up (TDU) episode permits partial mixing processes 
between material of the He$-$inter- shell
and the convective envelope.
During the TDU, few protons are assumed to penetrate into 
the top layers of the He$-$inter- shell, and subsequently react via 
$^{12}$C(p, $\gamma$)$^{13}$N($\beta^{+}$$\nu$)$^{13}$C chain, 
enriching in $^{13}$C a thin region at the top of the
He$-$intershell, the $^{13}$C-pocket. 
At $T$ $\sim$ 0.9$\times$10$^{8}$ K, $^{13}$C burns radiatively 
during the interpulse period \citep{straniero95}. 
The second neutron source
is marginally activated during the convective 
thermal pulses, when the maximum temperature reaches 
$T$ $>$ 2.5$\times$10$^{8}$ K. 
This maximum temperature slightly increases with pulse number
and decreasing the metallicity \citep[see][]{cristallo}.
Although this second neutron burst represents 
only a few percent of the total neutron exposure, it modifies 
the abundance patterns of several branchings that are 
sensitive to temperature and neutron density.

Observations of $s-$enhanced stars at various metallicities 
require a range of $s-$process efficiencies \citep{SCG08}.
Starting from the ST case\footnote{Our $^{13}$C-pocket extend 
in mass for 9.4 $\times$ 10$^{-4}$ $M_{\odot}$
(about 1/20 of the typical mass involved in a TP),
and contains 4.7 $\times$ 10$^{-6}$ $M_{\odot}$ of $^{13}$C 
(\textsl{ST} case).}, which has been shown 
to reproduce the solar main component for AGB models 
of half-solar metallicity \citep{arlandini99}, we consider 
a large range of $^{13}$C$-$pocket
efficiencies between ST/60 up to ST$\times$2.

The $s-$process is characterized by three abundance peaks,
the Zr-peak (light-s, ls), the Ba-peak (heavy-s, hs), and the Pb-peak 
at the termination of the $s-$process path, corresponding to
the magic neutron numbers $N$= 50, 82, 126.
We adopt the spectroscopic definition [ls/Fe] = log$_{10}$(ls/Fe)$_{\star}$ - 
log$_{10}$(ls/Fe)$_{\odot}$ and analogously for [hs/Fe] and [Pb/Fe].
We assume ls = (Y,Zr) and hs = (La,Nd,Sm), because Sr and 
Ba have few and saturated lines \citep[see][]{busso95}.
To characterize the whole $s-$process distribution,
two $s-$process indicators, [hs/ls] and [Pb/hs],
independent of the specific envelope abundance 
enhancement, are necessary.
The spectroscopic $s-$process abundances observed in CEMP(s) 
stars depend on the fraction of the AGB mass transferred
by stellar winds, whilst the $s-$process indicators [hs/ls] 
and [Pb/hs] remains unchanged.
We may introduce a dilution factor between the AGB mass 
transferred and the original envelope of the observed
star. We will define the logarithmic ratio `dil' as\\
$\rm{dil = \log\left(\frac{M^{env}_{\star}}{\Delta M^{trans}_{AGB}}\right)}$,
where $M^{env}_{\star}$ represents the mass of the convective 
envelope of the observed star before the mixing, 
and $\Delta M^{trans}_{AGB}$ is the AGB total mass transferred.
For instance, dil $\sim$ 0 dex means that the 
convective envelope mass of the observed star, which was interested
in the mixing processes, is of the same order of the $s-$rich material 
transferred from the AGB. 
Low-metallicity main sequence stars, with mass of
about 0.8 -- 0.9 $M_\odot$, are characterized by a
very thin convective subphotospheric layer 
($M$ $\leq$ 10$^{-3}$ $M_{\odot}$). 
Instead, for a giant that already suffered the first dredge$-$up (FDU) episode, 
the convective envelope extends to 8/10 of the mass of the star.
For subgiants, different degrees of FDU deepness can be reached,
and a range of dilution factors can be adopted.

For halo stars, an initial $\alpha-$enhancement of [$\alpha$/Fe] 
= 0.3 -- 0.5 dex is adopted for Mg, Si, Ca, and Ti. 
As for [O/Fe], we adopt a linear increase with decreasing
metallicity: [O/Fe] = $-$0.4$\times$[Fe/H] according to 
\citet{abia01}.
Furthermore, during the AGB phase, a primary contribution to $^{16}$O 
derives from the partial He$-$burning during the thermal pulses,
through $^{12}$C($\alpha$, $\gamma$)$^{16}$O.

Many CEMP(s) stars are also enriched in
$r-$process elements, CEMP(s+r).
About 70$\%$ of solar La is synthesized by the 
$s-$process \citep{winckler06}, whilst 94$\%$ of
solar Eu is provided by the $r-$process.
[La/Eu] is the major indicator of $s+r$ enrichment in stars.
The theoretical AGB predictions from a pure main $s-$process 
give [La/Eu]$_{s}$ $\sim$ 1.
The $r-$process elements are not synthesized by AGBs;
consequently in order to explain much lower observed [La/Eu] 
ratio (down to 0), different scenarios have been proposed in the 
literature\footnote{See \citet{jonsell06} and references therein.}.
\citet{vanhalacameron98} showed through numerical
simulations how Type II supernova ejecta may interact with a molecular 
cloud, polluting it with freshly synthesized material
(in particular $r-$process), likely triggering the formation 
of binary systems, which consists of stars with low mass.
We assume that the molecular cloud, from 
which the binary system formed, was pre$-$enriched in $r-$elements
because of supernova Type II pollution
\citep{bisterzo08FSIII,bisterzo08perugia,SCG08}. 
Our choice of the initial $r-$rich element abundances 
was made considering the $r-$process solar 
predictions from \citet{arlandini99}, taking into 
account the different $r-$process percentage to solar abundances 
that contributes to each isotope of a given element.

Two papers are in preparation that will describe accurately
the AGB theoretical models and interpret the abundances 
of all CEMP(s) and CEMP(s+r) stars discovered so far. 
The aim of this paper is to discuss five of these 
objects as specific examples:
the main$-$sequence/turnoff star CS 29526$-$110,
the subgiant CS 22880$-$074 (Sect.\ref{cemps}), and the three giants, 
CS 22942$-$019 (Sect.\ref{cemps}), LP 625$-$44, and HE 0202$-$2204. 
For HE 0202$-$2204 no lead is measured and we provide a 
theoretical prediction (Sect.\ref{leadpredictions}). 
Two of the five stars, CS 29526$-$110 and LP 625$-$44,
are CEMP(s+r) (Sect.\ref{cemps+r}).

\section{CEMP(s) stars}\label{cemps}

The subgiant \textbf{CS 22880$-$074} ({\textit T}$_{\rm eff}$ = 5850 K and 
log $g$ = 3.8; [Fe/H] = $-$1.93) analyzed by 
\citet{aoki02c,aoki02d}, with a mild $s-$process enhancement:
[ls/Fe] $\sim$ 0.3, [hs/Fe] $\sim$ 1.2, and [Pb/Fe] $\sim$ 1.9.
\citet{aoki07} reported a subsolar [Na/Fe], including 
NLTE corrections ($-$0.7 dex for this object).
The ratios [hs/ls] and [Pb/hs] observed give an indication of the
whole $s-$process distribution.
The difference between the predicted [La/Fe] by an AGB model and 
the one observed in CS 22880$-$074 gives a first assessment 
of the dilution of the AGB material, afterward optimized with 
a more careful analysis of the uncertainties of the singles species. 
The goodness of the fit is tested for different AGB masses by 
considering all the elements from carbon to lead, weighing the 
number of lines detected for each of them.
The lower initial mass modeled ($M^{\rm AGB}_{\rm ini}$ = 1.2 $M_{\odot}$
with only three thermal pulses followed by TDU) and a case ST/9,
predicts [La/Fe]$_{\rm th}$ = 1.47 at [Fe/H]$_{\rm th}$ = $-$2: 
no solutions are possible for this star without a dilution factor 
because of the low $s-$enhancement.

\begin{figure}
\begin{center}
\includegraphics[scale=0.45, angle=-90]{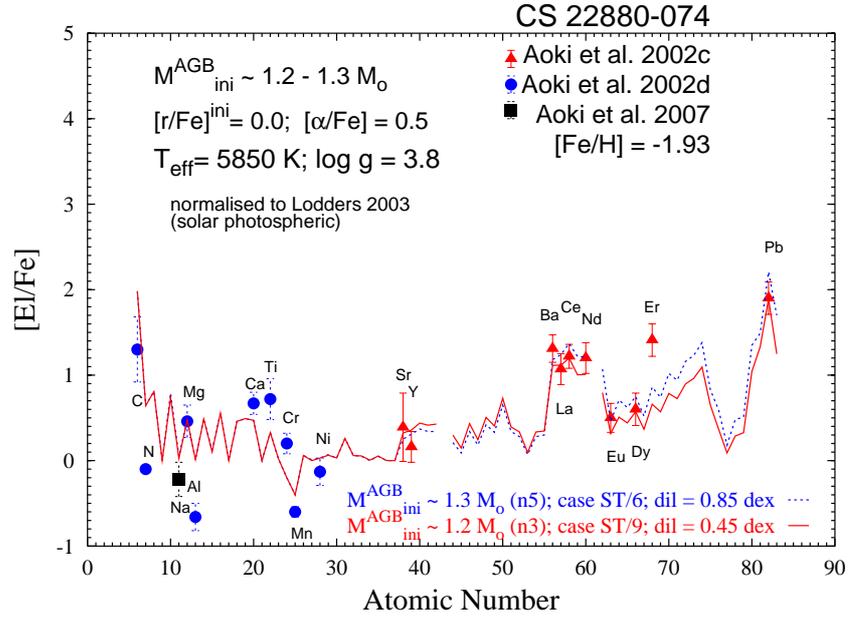}
\caption{Comparison of the [El/Fe] abundances of CS 
22880$-$074 by \citet{aoki02c,aoki02d,aoki07},
with AGB stellar models of 1.2 and 1.3 $M_{\odot}$, ST/6 and ST/9, and
dil = 0.45 and 0.85 dex, respectively. All the data and the theoretical expectations
are normalized to the solar photospheric abundances by \citet{lodders03}.}\label{CS074}
\end{center}
\end{figure}

We show in Fig.~\ref{CS074} the two theoretical interpretations 
for this star using $M^{\rm AGB}_{\rm ini}$ = 1.2 $M_{\odot}$, 
a case ST/9 and dil = 0.45 dex (red line), 
and $M^{\rm AGB}_{\rm ini}$ = 1.3 $M_{\odot}$,
a case ST/6 and dil = 0.85 dex (blue dotted line). 
The predicted [N/Fe] $\sim$ 0.8 by these models, takes into account 
the typical enhancement of $\sim$ 0.6 dex due to the FDU, and the
additional contribution from the H$-$shell ashes mixed with the envelope 
during the TDU.
This second contribution increases with the number of thermal pulses.
AGB models of higher initial mass ($M^{\rm AGB}_{\rm ini}$ = 1.5 
and 2 $M_{\odot}$, with dil = 1.75 and 2.2 dex, respectively)
would reproduce [hs/Fe] and [Pb/Fe], but
overestimated of $\sim$ 0.5 dex the observed [Y/Fe]
(no zirconium is detected in this star), and similarly for
[Na/Fe].
The observed [La/Eu] ratio is in agreement with a pure $s-$process
contribution ([r/Fe]$^{\rm ini}$ = 0, which corresponds to [La/Eu] 
= 0.8 for the case shown in Fig.~\ref{CS074}).
[Er/Fe] is about 0.5 dex higher than the other two $r-$process elements 
[Eu/Fe] and [Dy/Fe]\footnote{Note that for Eu, Dy and Er only one 
line is detected.}.
 
No radial velocity variations were measured for CS 22880$-$074
\citep{PS01,aoki02c}.
Preston (these Proceedings) confirms the non-detection of velocity
variations with his data over a period of 16 years. 
This is not a stringent argument against binarity 
\citep[see also][]{tsangarides05}. 

The giant \textbf{CS 22942$-$019} was analyzed by \citet{aoki02c,aoki02d} and
\citet{schuler08}.

\onecolumn
\begin{figure}
\begin{center}
\includegraphics[scale=0.45, angle=-90]{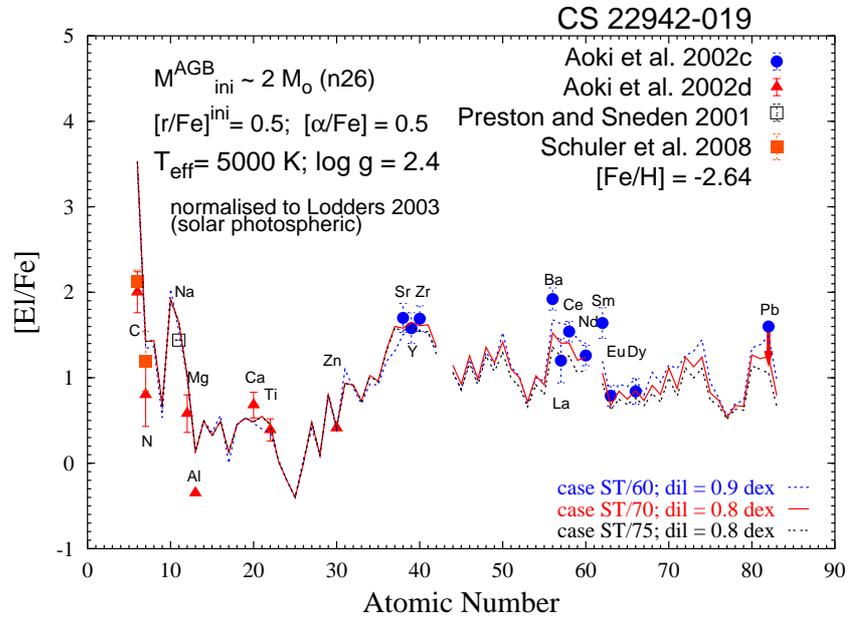}
\caption{Comparison of the [El/Fe] abundances of CS 22942$-$019
by \citet{aoki02c,aoki02d,aoki07} and \citet{schuler08} using 
 AGB stellar models of 2 $M_{\odot}$, 
ST/60, ST/70 and ST/75, and dil = 0.8 -- 0.9 dex.}\label{CS019}
\end{center}
\end{figure}
\twocolumn

\onecolumn
\begin{figure}
\begin{center}
\includegraphics[scale=0.45, angle=-90]{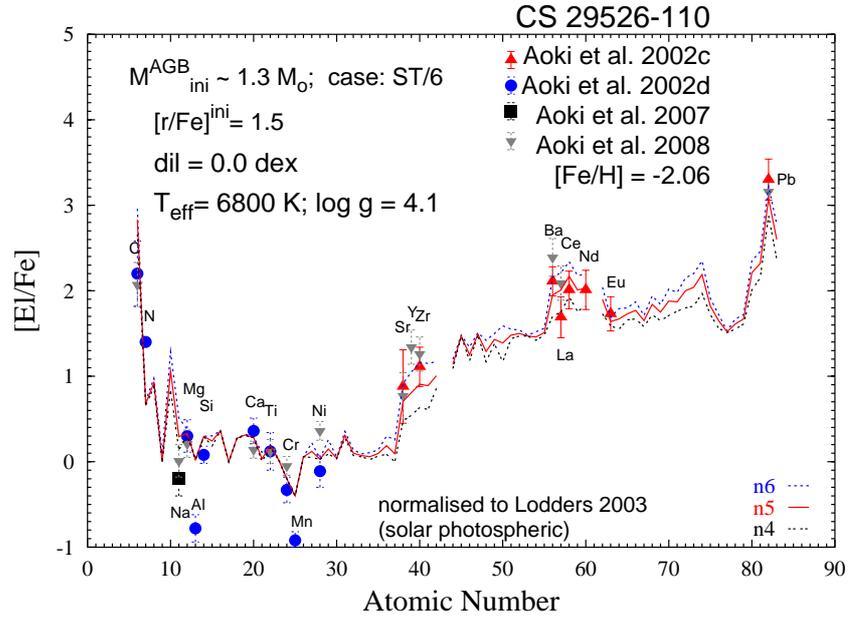}
\caption{Comparison of the [El/Fe] abundances of CS 
29526$-$110 by \citet{aoki02c,aoki02d,aoki07,aoki08},
with AGB stellar models of 1.3 $M_{\odot}$, ST/6, and
[r/Fe]$^{\rm ini}$ = 1.5. The envelope overabundances after the
fourth, fifth and sixth thermal pulse are 
shown (n4, n5, n6), corresponding to an uncertainty 
of the initial AGB mass ($\pm$ 0.05 $M_\odot$).}\label{CS110}
\end{center}
\end{figure}

\begin{figure}
\begin{center}
\includegraphics[scale=0.45, angle=-90]{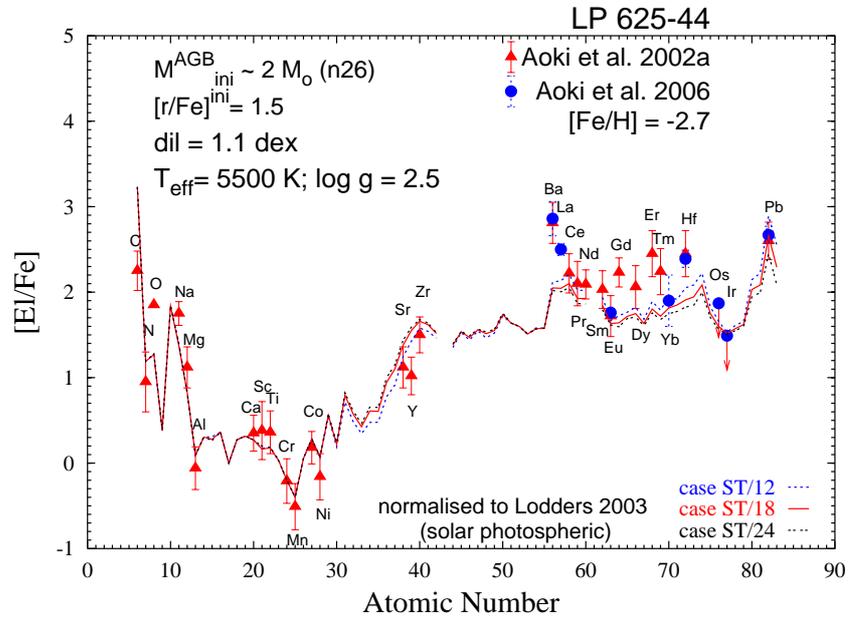}
\caption{Comparison of the [El/Fe] abundances of LP 625$-$44
by \citet{aoki02a,aoki06}, with AGB stellar models of 2 
$M_{\odot}$, ST/12, ST/18, ST/24, and dil = 1.1 dex. 
Note that the observed [O/Fe] is very uncertain in this star \citep[see][]{aoki02a}.}\label{LP44A}
\end{center}
\end{figure}

\begin{figure}
\begin{center}
\includegraphics[scale=0.45, angle=-90]{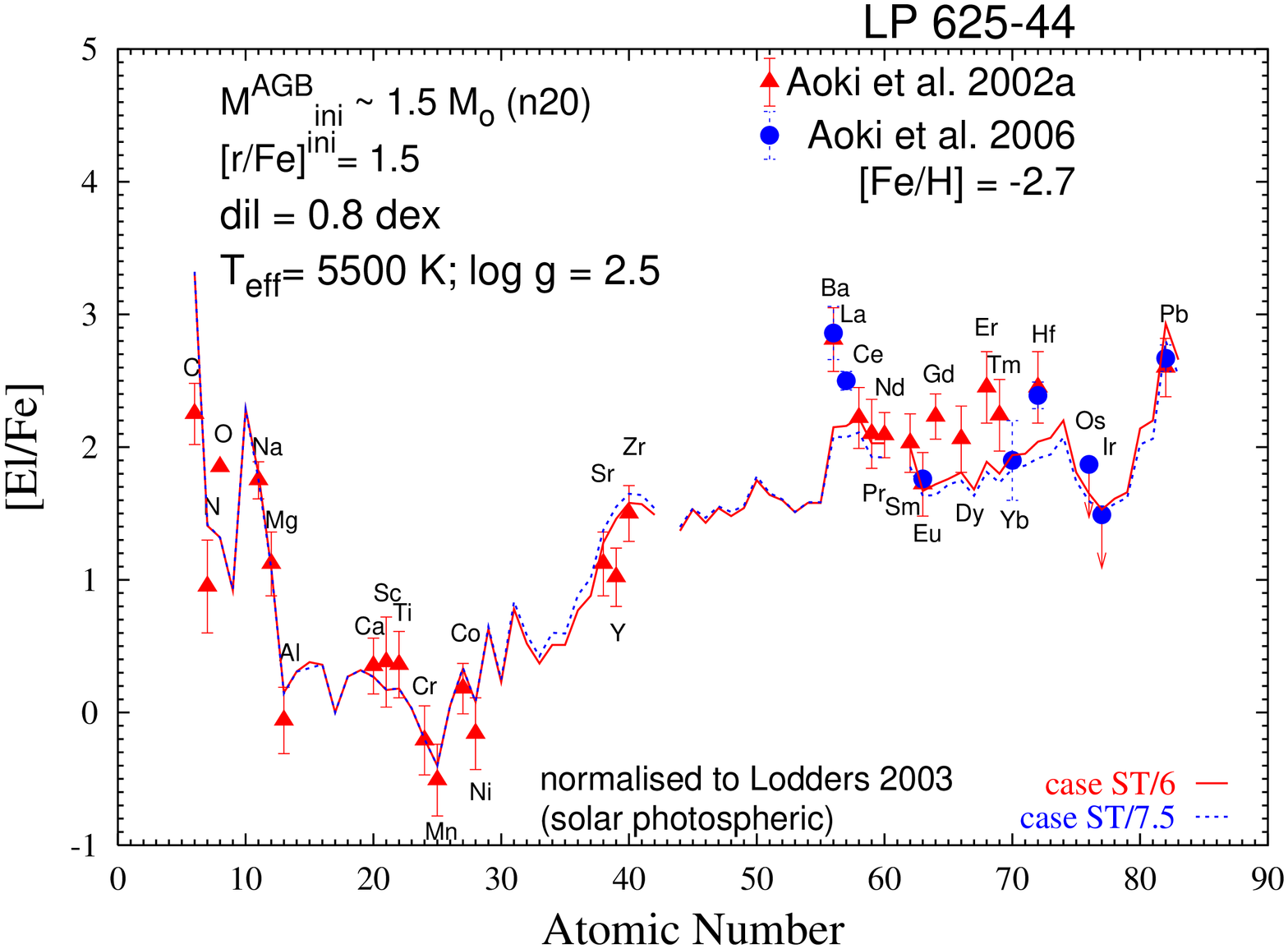}
\includegraphics[scale=0.45, angle=-90]{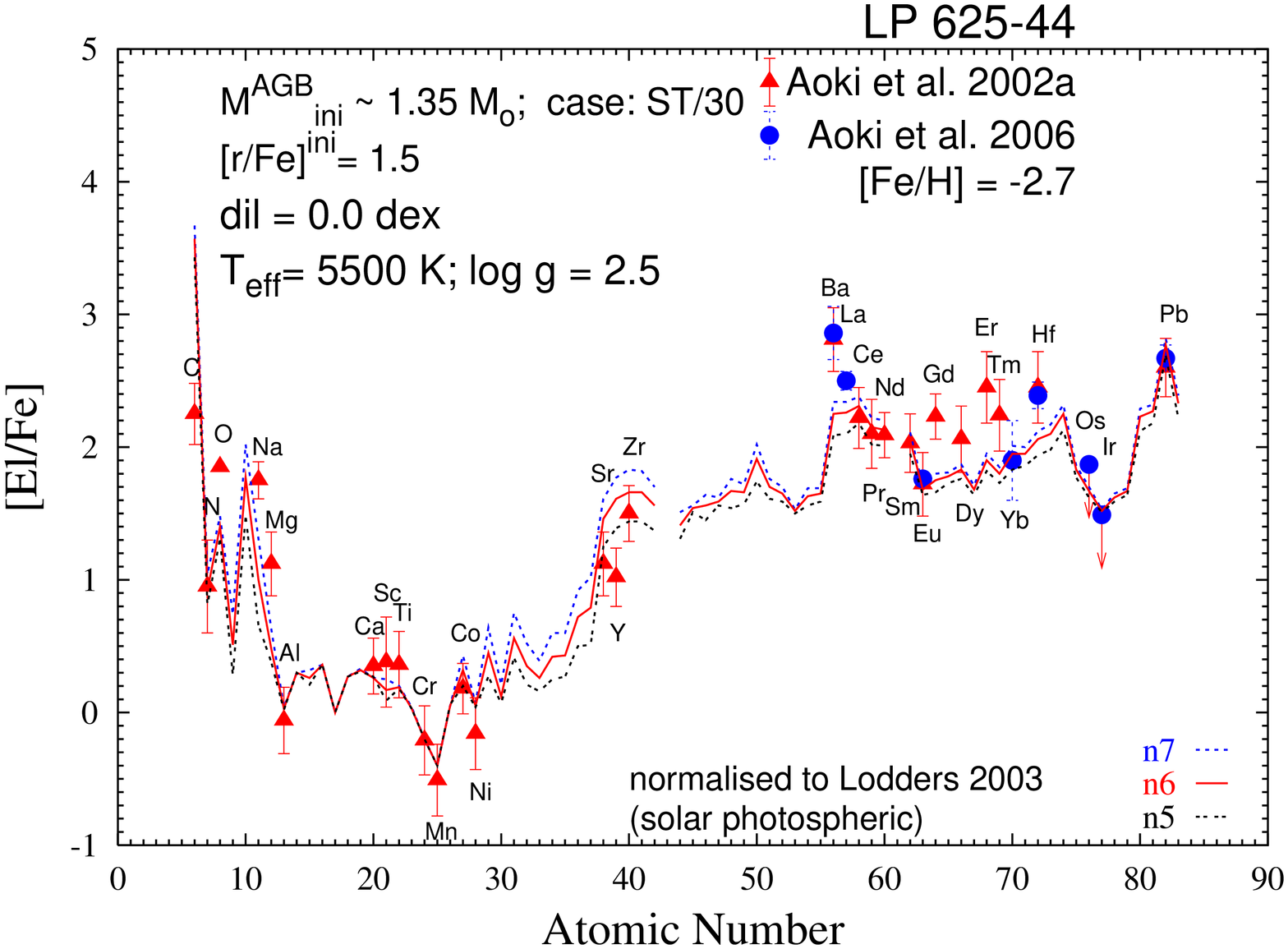}
\caption{The same as Fig.~\ref{LP44A}, but with AGB stellar models
1.5 $M_{\odot}$, ST/6, ST/7.5 and dil = 0.8 dex (top panel)
and 1.35 $M_{\odot}$, ST/30 and dil = 0.0 dex (bottom panel).}\label{LP44B}
\end{center}
\end{figure}
\twocolumn

\noindent \citet{PS01} derived its period, 505 d $\leq$ $P$ $\leq$ 3125 d,
discovering it is a long period binary.
A very high [Na/Fe] = 1.44 was found by \citet{PS01}.
For elements belonging to the hs peak, we expect differences in
the predictions at most of $\Delta$[hs/Fe] $\sim$ 0.2 dex, 
in disagreement with the observed [Ba/La] = 0.7 $\pm$ 0.2.
Ce and Nd abundances appear to be more reliable with 6 and
7 lines, while Ba, La and Sm have 3, 2 and 1 lines, respectively.
Only an upper limit for lead is available ([Pb/Fe] $\leq$ 1.6
and [Pb/hs] $\leq$ 0), suggesting AGB models with a 
low $s-$process efficiency. 
For $M^{\rm AGB}_{\rm ini}$ = 2 $M_{\odot}$ and a case ST/70 
(Fig.~\ref{CS019}), a dil = 0.8 -- 0.9 dex is applied, 
in agreement with a giant after the FDU episode. 
The high predicted [Na/Fe] is derived from $^{22}$Ne(n,$\gamma$)$^{23}$Na. 
Concerning [Mg/Fe], one has to take account of the initial enhancement 
of $^{24}$Mg and of the additional contribution by the
$^{23}$Na(n,$\gamma$)$^{24}$Mg reaction; moreover, a very high
overabundance of $^{25}$Mg and $^{26}$Mg is produced via 
$^{22}$Ne($\alpha$,n)$^{25}$Mg and $^{22}$Ne($\alpha$,$\gamma$)$^{26}$Mg,
and by neutron captures.
With low initial mass cases a lower dilution would be required
(dil = 0.5 dex for $M^{\rm AGB}_{\rm ini}$ = 1.5 $M_{\odot}$
and no dilution for $M^{\rm AGB}_{\rm ini}$ = 1.35 
$M_{\odot}$ model).
A higher mass model, $M^{\rm AGB}_{\rm ini}$ = 3 $M_{\odot}$,
has to be excluded, because to reproduce [hs/Fe] and [Pb/Fe] a dil 
$\sim$ 1.0 dex is required, whilst [ls/Fe] (as well as Na and Mg) 
would be overestimated.

\section{CEMP(s+r) stars}\label{cemps+r}

Among the objects with Eu detected, sixteen stars are
 CEMP(s+r)\footnote{We classified 
a stars as CEMP(s+r) if [r/Fe]$^{\rm ini}$ $\geq$ 1, where 
[r/Fe]$^{\rm ini}$ is the initial $r-$enhancement assumed 
to interpret the observations.} 
($\sim$ 40\%), five of them showing
a very high initial $r-$enrichment  
\citep[{[r/Fe]}$^{\rm ini}$ = 2.0; see][]{SCG08}.
In this Section, we discuss the two stars CS 29526$-$110 and
LP 625$-$44.

\textbf{CS 29526$-$110} has been analyzed several times 
\citep{aoki02c,aoki02d,aoki07,aoki08}. 
It is a single$-$lined binary \citep{aoki02d,tsangarides05}, 
although its period remains uncertain.
This object appears to be a turnoff star or a slightly evolved 
subgiant ($T_{\rm eff}$ = 6500 $\pm$ 200 K, log $g$ = 3.2 $\pm$ 0.5).
  
The theoretical interpretation shown in Fig.~\ref{CS110}
corresponds to an AGB model of $M^{\rm AGB}_{\rm ini}$ = 1.3 $M_{\odot}$, 
case ST/6 and no dilution factor. 
Nitrogen abundance is very uncertain, because of the 
difficulty of the CN band detection. 
The observed [Na/Fe] is about solar \citep{aoki07,aoki08}.
The most recent La measurement \citep{aoki08} better 
agrees with the prediction.
%
With a higher initial AGB mass models ($M^{\rm AGB}_{\rm ini}$ 
$\geq$ 1.4 $M_{\odot}$), Na and Mg are overestimated, whilst 
the $s-$process elements are would be reproduced with
a proper dilution and $^{13}$C$-$pocket case. 
The [La/Eu] $\sim$ 0.3 suggests an initial $r-$process 
enrichment [r/Fe]$^{\rm ini}$ = 1.5, while a pure 
$s-$process model would predict [La/Eu]$_{\rm th}$ = 0.85.

\citet{aoki02a} carried out a detailed analysis of 
the $s+r-$rich subgiant \textbf{LP 625$-$44}, subsequently
improved by determining the upper limits for two 
$r-$process elements, Os and Ir \citep{aoki06}. 
The binarity of this object was confirmed by radial 
velocity monitoring \citep{norris97,aoki00}, 
strongly supporting the mass transfer scenario, but 
the period has not been detected yet.
Very enhanced Na and Mg are observed in this star
([Na/Fe] = 1.75; [Mg/Fe] = 1.12). 
With $M^{\rm AGB}_{\rm ini}$ = 2 $M_{\odot}$ (Fig.~\ref{LP44A}), 
ST/18 and dil = 1.1 dex, we find a reasonable solution for 
all $s-$elements, and [Na/Fe] is acceptable within 2$\sigma$
uncertainty. 
Remember that Na is affected by the poorly 
understood corrections due to 3D atmospheric analysis and 
non$-$LTE calculation \citep{aoki08}.
The observed [Pb/hs] ratio is low ($\sim$ 0.35 dex), and a low
neutron flux (ST/18) is needed to interpret the whole $s-$process 
pattern.
Also for $M^{\rm AGB}_{\rm ini}$ = 1.5 $M_{\odot}$, ST/6 and ST/7.5, 
and dil = 0.8 dex, a satisfactory solution found (Fig.~\ref{LP44B}, 
top panel). 
In Fig.~\ref{LP44B}, bottom panel, we extend to the light elements
the solution already presented in \citet{aoki06} for 
a $M^{\rm AGB}_{\rm ini}$ = 1.35 $M_{\odot}$ model.
However, in this case [Na/Fe] and [Mg/Fe] would be 
underestimated.
[Y/Fe] is lower than the theoretical results -- note that we 
predict differences of at most the order of 0.2 dex between 
[Y/Fe] and [Zr/Fe] -- while [Ba/Fe] and [La/Fe] are both 
underestimated by the model. 
We considered Ce as more reliable among the second $s-$peak 
with 33 lines detected, while 4 and 13 lines are used for Ba
 and La, respectively. 
To match the $r-$process abundances, we used an initial 
$r-$enrichment [r/Fe]$^{\rm ini}$ = 1.5 dex. 
This choice is consistent with the upper limits of Os and Ir.

\section{Lead predictions}\label{leadpredictions}

For several stars among CEMP(s) and CEMP(s+r),
no lead is measured, and we give our theoretical 
predictions. In some cases, due to the uncertainty of the 
spectroscopic observations, we can only hypothesize a range
of expectations.

\begin{figure}
\begin{center}
\includegraphics[scale=0.45, angle=-90]{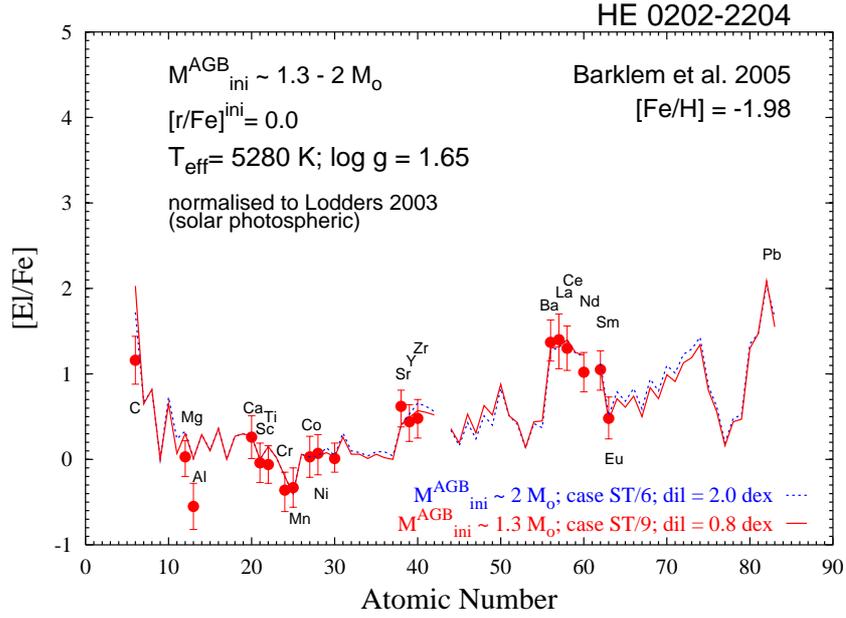}
\caption{Comparison of the [El/Fe] abundances of HE 0202$-$2204
by \citet{barklem05}, with a AGB stellar models
 of 1.3 and 2 $M_{\odot}$, ST/9 and ST/6, and dil = 0.8 and 2.0 dex,
 respectively. We predict [Pb/Fe]$_{\rm th}$ $\sim$ 2.1.
}\label{HE0202}
\end{center}
\end{figure}

As example we discuss here \textbf{HE 0202$-$2204}
studied by \citet{barklem05}. This giant can be interpreted with all
the initial masses in the range $M^{\rm AGB}_{\rm ini}$ = 1.3 and 
2 $M_{\odot}$, adopting low $^{13}$C$-$pockets (ST/9 and ST/6) 
and dil = 0.8 and 2.0 dex, respectively (Fig.~\ref{HE0202}).  
The lead predicted is [Pb/Fe]$_{\rm th}$ $\sim$ 2.1. 
All the $s-$process elements are well matched.

\section{Conclusions}

We analyzed the AGB model results for different initial masses
 ($M^{\rm AGB}_{\rm ini}$ = 1.3, 1.4, 1.5, 2 and 3 $M_{\odot}$), 
metallicities ($-$3.6 $\leq$ [Fe/H] $<$ $-$1) and $s-$process 
efficiencies (ST/150 $\leq$ $^{13}$C$-$pocket $\leq$ ST$\times$2), and
tested these models through a comparison between theoretical predictions
and spectroscopic abundances of the five CEMP(s) and CEMP(s+r) stars:
CS 22880$-$074, CS 22942$-$019, CS 29526$-$110, HE 0202$-$2204, 
and LP 625$-$44.
By comparing the [Na/Fe] (and [Mg/Fe]) observations with theoretical models, 
we can obtain an indicator of the initial AGB mass \citep{bisterzo06}. 
For instance, the low [Na/Fe] observed in CS 22880$-$074 and CS 29526$-$110
may be interpreted with $M^{\rm AGB}_{\rm ini}$ $\sim$ 1.3 
$M_{\odot}$ models.
However, a large uncertainty affects Na due to poorly 
understood NLTE effects and 

\onecolumn

\begin{table*}[h!]                   
\caption{Metallicity, atmospheric parameters, evolutionary stage
(MS means main$-$sequence star; SG means subgiant; G means giant), 
and some of the observed elements for the five stars analyzed.
In column~6, stars after the FDU are labeled as `yes', and viceversa `no'.}
\label{fivestar}                              
\begin{tabular}{ccccccccccc}                 
\hline \\[0.3ex]
Star&[Fe/H]&$T_{\rm eff}$&log $g$&Type&FDU&[Na/Fe] &[Mg/Fe] &[ls/Fe] &[hs/Fe] &[Pb/Fe] \\[0.5ex]
\hline
CS 22880$-$074 &-1.93&5850&3.8&SG&no &-0.09&0.46&0.16&1.14&1.90\\
CS 22942$-$019 &-2.64&5000&2.4&G &yes &1.44&0.58&1.64&1.37&$\leq$1.6\\
CS 29526$-$110 &-2.06&6800&4.1&MS&no &-0.07&0.22&1.11&1.85&3.30\\
HE 0202$-$2204 &-1.98&5280&1.65&G&yes &-  &-0.01&0.44&1.14&-\\
LP 625$-$44    &-2.70&5500&2.5&G &yes &1.75&1.12&1.26&2.21&2.67\\
\hline
\end{tabular}                               
\end{table*} 

\begin{table*}[h!]                    
\caption{Summary of the theoretical interpretations for the five stars analyzed.} 
\label{fivestar1}                           
\begin{tabular}{cccccc}  
\hline \\[0.3ex]
Star&$M^{\rm AGB}_{\rm ini}$&$^{13}$C$-$pocket&dil &[r/Fe]$^{\rm ini}$ &Fig.\\[0.5ex]
\hline 
CS 22880$-$074 &1.2  & ST/9  & 0.5  & 0.0 &\ref{CS074}\\
-              &1.3  & ST/6  & 0.8  & 0.0 &\ref{CS074}\\
CS 22942$-$019 &1.35 & ST/75 & 0.0  & 0.5 & - \\
-              &2    & ST/70 & 0.8  & 0.5 &\ref{CS019}\\
CS 29526$-$110 &1.3  & ST/6  & 0.0  & 1.5 &\ref{CS110}\\
HE 0202$-$2204 &1.3  & ST/9  & 0.8  & 0.0 &\ref{HE0202}\\
-              &1.5  & ST/4  & 1.6  & 0.0 & -\\
-              &2    & ST/6  & 2.0  & 0.0 &\ref{HE0202}\\
LP 625$-$44    &1.35 & ST/30 & 0.0  & 1.5 &\ref{LP44B}, top panel\\
-              &1.5  & ST/6  & 0.8  & 1.5 &\ref{LP44B}, bottom panel\\
-              &2    & ST/18 & 1.1  & 1.5 &\ref{LP44A}\\
\hline
\end{tabular}                              
\end{table*} 

\twocolumn

\noindent for 3D atmospheric models. 
Another indicator of the initial AGB mass is the ls peak,  
since the [ls/Fe] ratio is more sensitive to the thermal pulse number.
Solutions in the range of 1.3 $\leq$ $M^{\rm AGB}_{\rm ini}/M_{\odot}$
$\leq$ 2 are accepted.

For CEMP(s+r) stars, we suggest that the molecular cloud, 
from which the binary system was formed, was pre$-$enriched in 
$r-$elements because of pollution by Type II supernovae. 
Different initial $r-$process enrichments are adopted to explain the
observations.
The two CEMP(s+r) stars discussed here, CS 29526$-$110
and LP 625$-$44, need an initial $r-$enrichment of
[r/Fe]$^{\rm ini}$ = 1.5.

The discrepancies between observed and predicted C and N 
can be explained by efficient Cool Bottom Processing (CBP), a mixing 
process which occurs in low$-$mass stars
\citep{nollett03,wasserburg06}. This process may decrease significantly
the C abundance in the envelope, while N increases. 

In Table~\ref{fivestar} the major observational characteristics
of the five stars discussed in this work are reported, and
in Table~\ref{fivestar1} the corresponding theoretical 
interpretation are presented.

Two forthcoming papers will extend the analysis to the
full sample of 85 CEMP(s) stars, with a detailed presentation of the
AGB stellar models, including intermediate mass AGBs,
and the data tables of the theoretical results with metallicity
and $s-$process efficiency changes.

A final consideration has to be mentioned.
For [Fe/H] $\leq$ $-$2.5 and mass $M^{\rm AGB}_{\rm ini}$ 
$\leq$ 1.5 $M_{\odot}$, a peculiar phenomenon has been 
advanced: an anomalous proton ingestion episode (PIE), 
from the envelope down to the convective He$-$intershell, 
occurs during the first TDU
 \citep[see][and references therein]{cristallo}.
The consequence is a huge TDU episode affecting 
the $s-$process distribution.

%

\section*{Acknowledgments} 

We thank the anonymous Referee for insightful comments,
which have helped improving the paper.
Work supported by the Italian MIUR$-$PRIN 2006 Project 
`Late Phases of Stellar Evolution: Nucleosynthesis in Supernovae,
AGB Stars, Planetary Nebulae.'


\end{document}